\documentclass[pre,twocolumn,nofootinbib]{revtex4}
\usepackage{amsmath,amssymb}

\usepackage{graphicx}
\usepackage{dcolumn}
\usepackage{bm}
\usepackage{subeqnar}
\usepackage{epstopdf,overpic,color,rotating}

\graphicspath{{figures/}}

\begin{document}
\preprint{APS/123-QED}

\title{Nonlinear Model Reduction for Complex Systems using Sparse Optimal Sensor Locations from Learned Nonlinear Libraries}
\author{Syuzanna Sargsyan*, Steven L. Brunton$^\dag$ and J. Nathan Kutz*}
\address{*Department of Applied Mathematics, University of Washington, Seattle, WA 98195-3925\\
$^\dag$Department of Mechanical Engineering, University of Washington, Seattle, WA 98195}
\date{\today}

\begin{abstract}
{We demonstrate the synthesis of sparse sampling and machine learning to characterize and model complex, nonlinear dynamical systems over a range of bifurcation parameters.  First, we construct modal libraries using the classical proper orthogonal decomposition to uncover dominant low-rank coherent structures.  }
Here, nonlinear libraries are also constructed in order to take advantage of the discrete empirical interpolation method and projection that allows for the approximation of
nonlinear terms in a low-dimensional way.  The selected sampling points are shown to be nearly optimal sensing locations for characterizing the underlying dynamics, stability, and bifurcations of complex systems.  
{The use of empirical interpolation points and sparse representation facilitate a family of local reduced-order models for each physical regime, rather than a higher-order global model, which has the benefit of physical interpretability of energy transfer between coherent structures.}
In particular, the discrete interpolation points and nonlinear modal libraries are used for sparse representation to classify the dynamic bifurcation regime in the complex Ginzburg-Landau equation.  It is shown that nonlinear point measurements are more effective than linear measurements when sensor noise is present.
\end{abstract}

\pacs{05.45.-a, 74.20.De, 05.45.Yv}
\maketitle

\section{Introduction}

The theoretical study of complex systems pervades the physical, biological and engineering sciences.  Today, these
studies are driven increasingly by computational simulations that are of growing complexity and dimension due to
numerical discretization schemes.  Yet most dynamics of interest are known ultimately to be low-dimensional
in nature~\cite{Cross:1993}, thus contrasting, and in antithesis to, the high-dimensional nature of scientific computing.
Reduced order models (ROMs) are of growing importance in scientific applications and computing as they help
reduce the computational complexity and time needed to solve large-scale, complex systems~\cite{rom_book}.
Specifically, ROMs provide a principled approach to approximating high-dimensional spatio-temporal systems, typically generated from
numerical discretization, by low-dimensional subspaces that produce nearly identical input/output characteristics of
the underlying nonlinear dynamical system.  However, despite the significant reduction in dimensionality, the complexity of evaluating higher-order nonlinear terms may remain as challenging as that of the original problem~\cite{eim,deim}.  The empirical interpolation method (EIM),  and the simplified discrete empirical 
interpolation method (DEIM) for the proper orthogonal decomposition (POD)~\cite{Lumley:1970,HLBR_turb}, overcome this difficulty by providing a computationally efficient method for discretely (sparsely) sampling and evaluating the nonlinearity.  These methods ensure that the computational complexity of ROMs scale favorably with the rank of the approximation, even with complex nonlinearities.  

An alternative computational strategy for handling the nonlinearity is based upon machine learning techniques whereby libraries of {\em learned} POD modes can be constructed and inner products pre-computed for a number of distinct dynamical regimes of the complex system~\cite{siads,Bright:2013,epj,Kutz:2013}.  This strategy also evokes the power of compressive sensing for efficiently identifying the active POD subspace necessary for a low-dimensional Galerkin-POD truncation~\cite{Lumley:1970,HLBR_turb}.
In this manuscript, we combine the power of the DEIM with the library building strategy.  Specifically, we show that
building libraries that encode the nonlinearities allows one to (i) take advantage of DEIM to evaluate the nonlinearities, (ii)
more robustly classify the dynamical regime the system	 is in, and (iii) identify the discrete and optimal sensor locations to evaluate
a nonlinear model reduction.  We demonstrate the full integration of the methods on a canonical model of
mathematical physics and nonlinear science, the cubic-quintic Ginzburg-Landau (CQGLE) equation.  

\subsection{Dimensionality Reduction}
Although a variety of dimensionality-reduction techniques exist, the ROM methodology considered here is based upon the proper 
orthogonal decomposition~\cite{Lumley:1970,HLBR_turb}.  The POD method is ubiquitous in the dimensionality 
reduction of physical systems.  It is alternatively referred to as principal components analysis (PCA)~\cite{Pearson:1901}, the 
Karhunen--Lo\`eve (KL) decomposition, empirical orthogonal functions (EOF)~\cite{lorenzMITTR56}, or the Hotelling 
transform~\cite{hotellingJEdPsy33_1,hotellingJEdPsy33_2}.   Snapshots (measurements) of many complex system often
exhibit low-dimensional phenomena~\cite{Cross:1993}, so that the majority of variance/energy is contained in a few modes computed from a singular
value decomposition (SVD).  For such a case, the POD basis is typically truncated at a pre-determined cut-off value, such as when the modal basis contain $99\%$ of the variance, so that only the first $r$-modes ($r$-rank truncation) are kept.  { There are numerous additional criteria for the truncation cut-off, and recent results derive a hard-threshold value for truncation that is optimal for systems with well-characterized noise~\cite{Gavish2014arxiv}.}  The SVD acts as a filter, and so often the truncated modes correspond to random fluctuations and disturbances.  If the data considered is generated by a dynamical system (nonlinear system of ordinary differential equations of order $n$), it is then possible to substitute the truncated POD expansion into the governing equation and obtain Galerkin projected dynamics on the rank-$r$ basis modes~\cite{HLBR_turb,Kutz:2013}.  Recall that we are assuming that the complex systems under consideration exhibit low-dimensional attractors, thus the Galerkin truncation with only a few modes should provide an accurate prediction of the evolution of the system.  Note that it has also been shown recently that it is possible to obtain a \emph{sketched}-SVD by randomly projecting the data initially and then computing the SVD~\cite{Fowler:2009,Gilbert:2012,Qi:2012}.

\subsection{Sparse Sampling}
EIM has been developed for the purpose of efficiently managing the computation of the nonlinearity in dimensionality reduction schemes, with DEIM specifically tailored to POD with Galerkin projection.  Indeed, DEIM approximates the nonlinearity by using a small, discrete sampling of points that are determined in an algorithmic way.  This ensures that the computational cost of evaluating the nonlinearity remains proportional to the rank of the reduced POD basis.   As an example, consider the case of an $r$-mode POD-Galerkin truncation. A simple cubic nonlinearity requires that the POD-Galerkin approximation be cubed, resulting in $r^3$ operations to evaluate the nonlinear term.  DEIM approximates the cubic nonlinearity by using $O(r)$ discrete sample points of the nonlinearity, thus preserving a low-dimensional ($O(r)$) computation, as desired.  The DEIM approach combines projection with interpolation.  Specifically, DEIM uses selected interpolation indices  to specify  an interpolation-based projection for a nearly optimal $\ell_2$ subspace approximating the nonlinearity.  EIM/DEIM are not the only methods developed to reduce
the complexity of evaluating nonlinear terms, see for instance the missing point estimation (MPE)~\cite{mpe} or gappy POD~\cite{gap1,gap2,Carlberg:2013} methods.  However, they have been successful
in a large number of diverse applications and models~\cite{deim}.  In any case, the MPE, gappy POD, and EIM/DEIM use a small selected set of spatial grid points to avoid evaluation of the expensive inner products required to evaluate nonlinear terms.

The discrete sampling points given by DEIM to evaluate the nonlinearity get a new interpretation in the current work.
Specifically, we show them to be the nearly optimal locations for placing sensors in the complex system in order to
(i) determine the dynamic regime of the system, (ii) reconstruct the current state of the system, and (iii) produce a
POD-Galerkin prediction (nonlinear model reduction) of the future state of the system.  Such tasks are accomplished by using ideas of sparse representation~\cite{Wright:2009} and compressive sensing~\cite{Donoho:2006,Donoho2,Candes:2006,Candes:2006a,Candes:2006c,Candes:2006b,Baraniuk:2007,Baraniuk:2009}.   In particular, the theory of compressive sensing shows that a small number of measurements are sufficient to perform a reconstruction provided there exists a sparse representation (or basis) of the data. Sparsity techniques have also been shown to be highly
effective for numerical solution schemes~\cite{Schaeffer:2013,Mackey:2014}.  In our case, the sparse basis is generated from a library learning procedure.  More than that, however, we also build libraries of the 
{\em nonlinearities}, thus pre-computing the low-dimensional structures observed in the different dynamical states of the complex system.
This allows for more robust dynamical classification as well as allowing easy evaluation of the nonlinear terms through DEIM.
The combination of library building, compressive sensing and DEIM is demonstrated to be a highly effective and intuitively appealing 
methodology for scientific computing applications.  It further highlights the need in modern scientific computing of complex systems to integrate a variety of data-driven modeling strategies, many of which are being developed under the aegis of machine learning, in order to most efficiently simulate large-scale systems.

\subsection{Physical Interpretation}
The ideas presented here are more than just numerical efficiencies.  Indeed, the methodology
identifies the underlying modal structures that drive the dynamics of the complex system, thus helping
to understand the fundamental interactions and physics of the system.   Throughout the development of 
20th-century physics and engineering sciences, the understanding of
many canonical problems has been driven by recasting the problem into its {\em natural} basis (mode) set.
The majority of classical problems from mathematical physics are linear Sturm-Liouville problems whose 
ideal modal representations are generated from eigenfunction decompositions, i.e. special functions.  In quantum mechanics, for instance, 
Gauss-Hermite (denoted by $H_n(x)$) polynomials are the natural basis elements for understanding the harmonic oscillator.   Likewise,
spherical harmonics (denoted by $Y_{l}^{m}(\theta,\varphi)$) are critical in the computation of atomic orbital electron configurations
as well as in representation of gravitational fields, the magnetic fields of planetary bodies and stars, 
and characterization of the cosmic microwave background radiation.  

For modern complex systems, nonlinearity plays
a dominant role and shapes the underlying modes, thus necessitating a new approach, such as that presented here, for extracting these 
critical spatio-temporal structures.  Remarkably, although nonlinearity creates new modal structures, it does not destroy the underlying
low-dimensional nature of the dynamics.  
{Distinct physical regimes may be obtained by varying bifurcation parameters, and these regimes will typically have different local bases and physical interactions.  Instead of developing a global interpolated model, which may obscure these distinct physical mechanisms, we advocate a hierarchy of models along with sparse sampling and machine learning to classify and characterize the system parameters from a few online measurements.  }
Methods that take advantage of such underlying structure are critical for 
developing theoretical understanding and garnering insight into the fundamental interactions of the physical, engineering and biological systems
under consideration.

The paper is outlined as follows.  In Sec.~\ref{sec:background}, an overview of the mathematical framework of the POD method and the DEIM is given.
This is followed up in Sec.~\ref{sec:model} with an introduction of the nonlinear dynamical system, i.e. the cubic-quintic Ginzburg-Landau equation, 
where the methods proposed here will be applied.  The library building procedure that encodes the various dynamical regimes of our model
equation are discussed in Sec.~\ref{sec:libs}.  Once the libraries are constructed, DEIM points, or sensor locations, are computed
in Sec.~\ref{sec:deim} and their ability to classify dynamical regimes is evaluated in Sec.~\ref{sec:class}.  The reconstruction of the dynamics
and future state projection is discussed in Sec.~\ref{sec:recon}.   A summary of our findings and an outlook on
the method is given in the concluding Sec.~\ref{sec:conclusion}.

\section{Background for Model Reduction}\label{sec:background}

Our innovations are built upon two key methods which are used for model reduction and
approximating nonlinear dynamical systems.  
The first approach is the well-known POD-Galerkin method, which is used to 
reduce the dimension of systems in a principled way.  
However, computing the form of the nonlinearity in the reduced-order system is an expensive offline computation, as inner products of the full high-dimensional system must still be computed.  Online evaluation of the nonlinear terms in the reduced order model may remain expensive, as these typically involve dense matrix or tensor operations of the same order as the degree of nonlinearity.  
The second approach highlighted is the DEIM algorithm~\cite{deim} which
reduces the complexity of evaluating the nonlinear terms.  In particular,
it gives a principled way to sparsely sample the nonlinearity in order to approximate
the nonlinear terms in a low-dimensional way.

\subsection{POD}\label{sec:back:pod}
Consider a  high-dimensional system of nonlinear differential equations that can arise, for example, from the finite difference 
discretization of a partial differential equation:
 \begin{equation}
 \label{eq:complex}
   \frac{d {\bf u}(t)}{dt}=L{\bf u}(t)+N({\bf u}(t)),
 \end{equation}
 where ${\bf u}(t)=[u_1(t) \,\, u_2(t) \,\, \cdots \,\, u_n(t)]^T\in \mathbb{R}^n$ and $n\gg 1$.  Typically under discretization of
 a single spatial variable, $u_j (t) = u(x_j, t)$ is the value of the field of interest at the spatial location $x_j$.
 The linear part of the dynamics is given by $L\in \mathbb{R}^{n\times n}$ and the nonlinear terms are
 in the vector $N({\bf u}(t)) = [N_1({\bf u}(t)) \quad N_2({\bf u}(t)) \quad \cdots \quad N_n({\bf u}(t))]^T\in \mathbb{R}^n$. 
 The nonlinear function is evaluated component-wise at the $n$ spatial grid points used for 
 discretization. 
 
 \begin{table*}[t]
   \caption{ \label{table:alg} DEIM algorithm for finding approximation basis for the nonlinearity and its interpolation indices.}
\begin{center}
\begin{tabular}[c]{|p{8cm}|p{8cm}|}
 \hline
  \multicolumn{2}{ |c| } { \bf  DEIM algorithm}\\  \hline
  \multicolumn{2}{ |c| } {\bf  {\color{blue}  Basis}} \\  \hline
  $\bullet$ collect data, construct snapshot matrix & $\mathbf{X}=[\mathbf{u}(t_1) \,\, \mathbf{u}(t_2) \,\, \cdots \,\, \mathbf{u}(t_p)]$\\  \hline
  $\bullet$ construct nonlinear snapshot matrix & $\mathbf{N}=[N({\bf u}(t_1)) \,\, N({\bf u}(t_2)) \,\, \cdots \,\, N({\bf u}(t_p))]$\\  \hline
  $\bullet$ singular value decomposition of $\mathbf{N}$ & $\mathbf{N}={\bf \Xi}{\bf \Sigma}_N {\bf W}_N^*$\\  \hline
  $\bullet$ construct approximating basis (first $m$ columns) & ${\bf \Xi}_m=[{\boldsymbol \xi}_1 \,\, {\boldsymbol \xi}_2 \,\, \cdots \,\, {\boldsymbol \xi}_m]$\\ \hline
  \multicolumn{2}{ |c| } {\bf  { \color{blue} Interpolation Indices (Iteration Loop)}} \\  \hline
  $\bullet$ choose the first index (initialization) 
  & $[\rho, \gamma_1]=\max|{\boldsymbol \xi}_1|$\\ \hline
  $\bullet$ approximate ${\boldsymbol \xi}_j$ by ${\boldsymbol \xi}_1,...,{\boldsymbol \xi}_{j-1}$  at indices $\gamma_1,...,\gamma_{j-1} $
  &  Solve for ${\bf c}$:
  ${\bf P}^T {\boldsymbol \xi}_j={\bf P}^T{\bf \Xi}_{j-1}{\bf c}$ with ${\bf P}=[{\bf e}_{\gamma_1} \,\, \cdots \,\, {\bf e}_{\gamma_{j-1}}]$  \\ \hline
  $\bullet$ select $\gamma_j$ and loop ($j=2, 3, ..., m$) & $[\rho,\gamma_j ]=\max|{\boldsymbol \xi}_j-{\bf \Xi}_{j-1} {\bf c}|$\\ \hline    
  \end{tabular}   
\end{center} 
 \end{table*}

 For achieving high accuracy solutions, $n$ is typically required to  
be a very large number, thus making the computation of the 
 solution expensive and/or intractable.  The POD-Galerkin method 
is a principled dimensionality-reduction scheme that approximates the function ${\bf u}(t)$ with rank-$r$
optimal basis functions where $r\ll n$.  
These optimal basis functions are computed from a singular value decomposition of a series
of temporal snapshots of the complex system.   Specifically, suppose snapshots of the
state, ${\bf u}(t_j)$ with $j=1, 2, \cdots , p$, are collected.  The snapshot matrix ${\bf X}=[{\bf u} (t_1) \,\, {\bf u}(t_2) \,\, \cdots \,\, {\bf u}(t_p)]
  \in \mathbb{R}^{n\times p}$ is constructed and the SVD of ${\bf X}$ is computed: 
  ${\bf X}={\bf \Phi\Sigma W}^*$.  The $r$-dimensional basis for optimally approximating 
 ${\bf u}(t)$ is given by the first $r$ columns of matrix ${\bf \Phi}$, denoted by ${\bf \Phi}_r$.
 Thus the POD-Galerkin approximation is given by 
 \begin{equation} {\bf u}(t)\approx {\bf \Phi}_r {\bf a}(t)
 \label{eq:podG}
 \end{equation} 
where ${\bf a}(t)\in \mathbb{R}^r$ is the time-dependent coefficient vector and $r\ll n$. 
 Plugging this modal expansion into the governing equation (\ref{eq:complex}) and applying orthogonality (multiplying
 by ${\bf \Phi}_r^T$) gives the dimensionally reduced evolution
   \begin{equation}
   \label{eq:pod}
     \frac{d{\bf a}(t)}{dt}={\bf \Phi}_r^T L {\bf \Phi}_r {\bf a}(t)+ {\bf \Phi}_r^T N({\bf \Phi}_r {\bf a}(t)).
   \end{equation}
By solving this system of much smaller dimension, the solution of a high-dimensional complex
system can be approximated.

This standard POD procedure~\cite{HLBR_turb} has been a ubiquitous algorithm in 
the reduced order modeling community.  However, it also helps illustrate the need for
innovations such as DEIM, Gappy POD and/or MPE.  Consider the nonlinear component
of the low-dimensional evolution (\ref{eq:pod}):  ${\bf \Phi}_r^T N({\bf \Phi}_r {\bf a}(t))$.
For a simple nonlinearity such as $N(u(x,t))=u(x,t)^3$, consider its impact on a spatially-discretized, two-mode
POD expansion: $u(x,t) = a_1(t) \phi_1(x) + a_2(t) \phi_2(x)$.  The
algorithm for computing the nonlinearity would require the evaluation:
\begin{equation}
  u(x,t)^3 = a_1^3 \phi_1^3   + 3 a_1^2 a_2 \phi_1^2 \phi_2 + 3 a_1 a_2^2 \phi_1 \phi_2^2  + a_2^3 \phi_2^3 \, .
\end{equation}
The dynamics of $a_1(t)$ and $a_2(t)$
would then be computed by projecting onto the low-dimensional basis set by taking
the inner product of this nonlinear term with respect to both $\phi_1$ and $\phi_2$.
Thus the number of computations not only doubles, but the inner products must be computed
with the $n$-dimensional vectors.  Methods such as DEIM overcome this high-dimensional
computation and instead produce an $O(r)$ dimensional evaluation of the nonlinear terms.

\subsection{DEIM}\label{sec:back:deim}
As outlined in the previous section, the shortcomings of the POD method are
generally due to the evaluation of the nonlinear term $N({\bf \Phi}_r {\bf a}(t))$. 
To avoid this difficulty, the DEIM approximates $\mathbf{N}=N({\bf \Phi}_r {\bf a}(t))$ through projection
and interpolation instead of evaluating it directly.  Specifically, a low-rank representation of the
nonlinearity is computed from the singular value decomposition
\begin{equation}
  {\bf N} = {\bf \Xi \Sigma}_N {\bf W}_N^*
\end{equation}
where the matrix ${\bf \Xi}$ contains the optimal (in an $\ell_2$ sense) basis set for spanning the nonlinearity.
Specifically, we consider the rank-$m$ basis set  ${\bf \Xi}_m=[\boldsymbol{\xi}_1 \,\, {\boldsymbol \xi}_2 \,\, \cdots \,\, {\boldsymbol \xi}_m]$
that approximates the nonlinear function ($m\ll n$ and $m\sim r$).  
The approximation to the nonlinearity  $\mathbf{N}$ is given by:
\begin{equation}
 \mathbf{N}\approx {\bf \Xi}_m {\bf c}(t)
\end{equation}
where ${\bf c}(t)$ is similar to ${\bf a}(t)$ in (\ref{eq:podG}).
Since this is a highly overdetermined system, a suitable vector ${\bf c}(t)$ can be found by selecting only $m$ 
rows of the system.  The DEIM algorithm was specifically developed to identify which
$m$ rows to evaluate.

The DEIM algorithm begins by considering the vectors ${\bf e}_{\gamma_j}\in\mathbf{R}^n$ which are the $\gamma_j$-th column of 
the $n$ dimensional identity matrix.  We can then construct the projection matrix 
${\bf P}=[{\bf e}_{\gamma_1} \,\, {\bf  e}_{\gamma_2} \,\, \cdots \,\, {\bf e}_{\gamma_m}]$ which is chosen so that ${\bf P}^T{\bf \Xi}_m$ is nonsingular. 
Then ${\bf c}(t)$ is uniquely defined from ${\bf P}^T\mathbf{N} ={\bf P}^T {\bf \Xi}_m {\bf c}(t)$, and thus, 
\begin{equation}
\label{eq:nonlinearity}
\mathbf{N}\approx {\bf \Xi}_m ({\bf P}^T {\bf \Xi}_m)^{-1} {\bf P}^T\mathbf{N}. 
\end{equation}
The tremendous advantage of this result for nonlinear model reduction is that the term ${\bf P}^T\mathbf{N}$   
 requires evaluation of nonlinearity only at $m$ indices, where $m\ll n$. 
The DEIM further proposes a principled method for choosing the basis vectors ${\boldsymbol \xi}_j$ and 
indices $\gamma_j$.  The DEIM algorithm, which is based upon a greedy-like search, is detailed in~\cite{deim} and
further demonstrated in Table~\ref{table:alg}.

\subsection{Application to ROMs}

POD and DEIM provide a number of advantages for nonlinear model reduction of complex systems.
POD provides a principled way to construct an $r$-dimensional 
subspace ${\bf \Phi}_r$ characterizing the dynamics.  DEIM
augments the POD method by providing a method to evaluate the 
problematic nonlinear terms using an $m$-dimensional subspace ${\bf \Xi}_m$ that represents
the nonlinearity.  Thus a small number of points, specifically $m$, can
be sampled to approximate the nonlinear terms in the ROM.

The method proposed here capitalizes on these methods by building
low-dimensional libraries associated with the full complex system dynamics
as well as the specific nonlinearities.  Moreover, the sparse measurement
locations computed by DEIM are found to be nearly optimal for sensor
placement.  Such sensors, as will be shown in what follows, can be used with sparse representation and
compressive sensing to (i) identify dynamical regimes, (ii) reconstruct
the full state of the system, and (iii) provide an efficient nonlinear model reduction and POD-Galerkin prediction
for the future state.  Moreover, we show that nonlinear measurements of the dynamical system
can be much more robust to noise for accomplishing the above tasks.

\section{Model Problem}\label{sec:model}
One of the canonical nonlinear PDEs in mathematical physics and pattern forming systems is the Ginzburg-Landau (GL) equation and its many-variants~\cite{Cross:1993}.  It has been used to model a variety of physical systems from condensed matter to biological waves.  Here we consider a variant of the GL equation arising in mode-locked laser theory that has cubic and quintic nonlinear terms  and a fourth-order derivative~\cite{kutz:SIAM}:
 \begin{eqnarray}
 && \hspace{-.2in} i {U}_t + \left(  \frac{1}{2} - i\tau  \right) {U}_{xx} - i \kappa {U}_{xxxx}
 + (1- i\mu) |{U}|^2 {U}  \nonumber \\
 &&  \hspace{.2in}+ (\nu - i \varepsilon) |{U}|^4 {U} - i\gamma {U} \! =\! 0,
  \label{eq:gl}
\end{eqnarray}
where ${U}(x, t)$ is a complex valued function of space and time.   Under discretization
of the spatial variable, $U(x,t)$ becomes a vector ${\bf u}$ with $n$ components, i.e. 
${\bf u}_j (t)=U(x_j,t)$ with $j=1, 2, \cdots n$.

\begin{figure}[tb]
\begin{overpic}[width=0.38\textwidth]{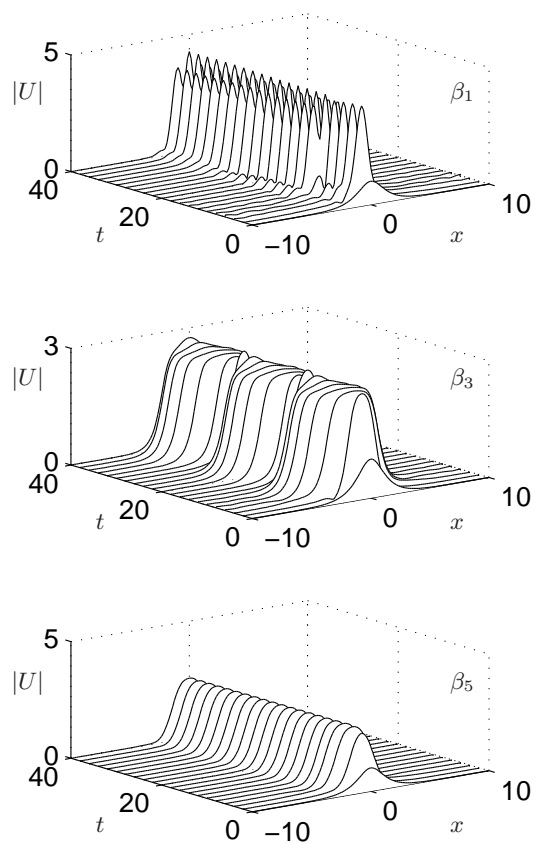}
\put(50,89){$\beta_1$}
\put(50,55){$\beta_3$}
\put(50,19){$\beta_5$}
\put(-2,19){$|U|$}
\put(-2,55){$|U|$}
\put(-2,89){$|U|$}
\put(8,2){$t$}
\put(8,38){$t$}
\put(8,72){$t$}
\put(50,2){$x$}
\put(50,38){$x$}
\put(50,72){$x$}
\end{overpic}
\caption{Evolution dynamics of (\ref{eq:gl}) for the parameter regimes $\beta_1$, $\beta_3$ and $\beta_5$ over the time interval $t\in[0,40]$.
The initial transients are quickly attenuated away, leaving the stable attractor for the given $\beta_j$ regime.  Sampling of the dynamics for
library building occurs once the transients have decayed.}
 \label{figure:3reg}
\end{figure}

An efficient and exponentially accurate numerical solution to (\ref{eq:gl}) can be found using standard spectral 
methods~\cite{Kutz:2013}.  Specifically, the equation is solved by Fourier transforming in the spatial
dimension and then time-stepping with an adaptive 4th-order Runge-Kutta method.  The extent of the spatial
domain is $x\in[-20,20]$ with $n=1024$ discretized points.   Note that in what follows, the indices for evaluation
of the nonlinear term correspond to the collocation points away from the center spatial point of the 
computational domain $x_{513}=0$.
Here,  we allow the parameters $\beta=(\tau, \kappa, \mu, \nu, \epsilon, \gamma )$ to vary 
in order to discover various dynamical regimes that exhibit low-rank structure and stable attractors.
Table \ref{table:vals} shows six different parameter regimes that have unique low-dimensional attractors (see \cite{epj}).
The evolution of the system for parameter regimes $\beta_1$, $\beta_3$ and $\beta_5$ is illustrated 
in Fig.~\ref{figure:3reg}.   Such stereotypical low-dimensional behaviors, which are commonly observed
in pattern forming systems~\cite{Cross:1993}, will serve as the basis for
our library building methodology, especially in regards to using a small number of measurements to
identify the $\beta_j$ regime, reconstruct the solution, and project a future state.  Although our results
are demonstrated on this specific PDE, the methodology is quite general.

\begin{table}[tb] 
\begin{center}
 \begin{tabular}[c]{|c|c|c|c|c|c|c|c|} 

\hline
&$\tau$&$\kappa$ &$\mu$ & $\nu$ & $\epsilon$ & $\gamma$ & description\\
\hline
$\beta_1$ & -0.3    &-0.05     &1.45    &0     &-0.1  &-0.5& 3-hump, localized\\
\hline
$\beta_2$ & -0.3    &-0.05     &1.4  &0   &-0.1  &-0.5  &  localized, side lobes\\
\hline
$\beta_3$ & 0.08    & 0      &0.66      &-0.1    &-0.1     &-0.1    &breather\\
\hline
$\beta_4$ & 0.125  &0    &1    &-0.6    &-0.1    &-0.1    & exploding soliton\\
\hline
$\beta_5$ & 0.08    &-0.05    &0.6    &-0.1    &-0.1    &-0.1    & fat soliton\\
\hline
$\beta_6$ & 0.08    &-0.05    &0.5    &-0.1    &-0.1    &-0.1    & dissipative soliton\\
\hline
   
 \end{tabular}
 \end{center}
 \caption{Values of the parameters from equation (\ref{eq:gl}) that lead to six distinct dynamical regimes. To exemplify our
 algorithm, the first, third and fifth regimes
 will be discussed in this paper.}
 \label{table:vals}
\end{table}

\begin{figure*}[t]
\begin{overpic}[width=1.0\textwidth]{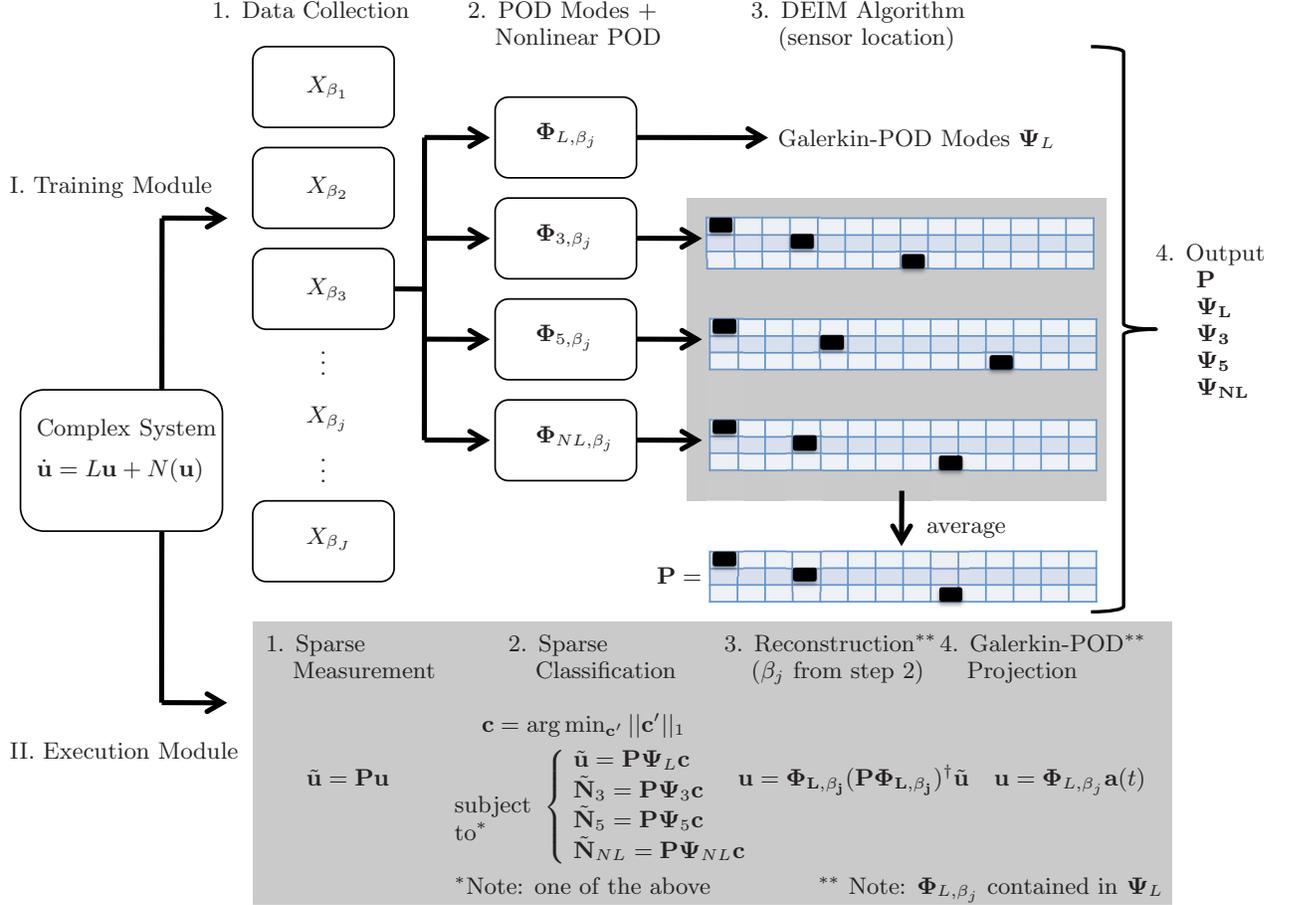}
\put(7,39){$\dot{\bf u}=L{\bf u}+N({\bf u})$}
\put(7,42){Complex System}
\put(5,60){I. Training Module}
\put(5,18){II. Execution Module}
\put(20,73){1.  Data Collection}
\put(39,73){2.  POD Modes +}
\put(41,71){Nonlinear POD}
\put(60,73){3.  DEIM Algorithm}
\put(62,71){(sensor location)}
\put(27,67.5){$X_{\beta_1}$}
\put(27,60){$X_{\beta_2}$}
\put(27,52.5){$X_{\beta_3}$}
\put(27,34){$X_{\beta_J}$}
\put(27,43){$X_{\beta_j}$}
\put(28,38.5){$\vdots$}
\put(28,46.5){$\vdots$}
\put(44,64){${\bf \Phi}_{L,\beta_j}$}
\put(44,56.5){${\bf \Phi}_{3,\beta_j}$}
\put(44,49){${\bf \Phi}_{5,\beta_j}$}
\put(44,41.5){${\bf \Phi}_{NL,\beta_j}$}
\put(62,63.5){Galerkin-POD Modes ${\bf \Psi}_{L}$}
\put(53,31){${\bf P}=$}
\put(73,34.8){average}
\put(90,55){4. Output}
\put(93,53){${\bf P}$}
\put(93,51){$\bf \Psi_L$}
\put(93,49){$\bf \Psi_3$}
\put(93,47){$\bf \Psi_5$}
\put(93,45){$\bf \Psi_{NL}$}
\put(27,16){$\tilde{\bf u}={\bf P} {\bf u}$}
\put(24,26){1.  Sparse}
\put(26,24){Measurement}
\put(42,26){2.  Sparse}
\put(44,24){Classification}
\put(58,26){3.  Reconstruction$^{**}$}
\put(60,24){($\beta_j$ from step 2)}
\put(74,26){4.  Galerkin-POD$^{**}$}
\put(76,24){Projection}
\put(40,20){${\bf c}=\text{arg}\min_{{\bf c}'} ||{\bf c}'||_1$}
\put(38,14){$\text{subject} \,\, \left\{ \begin{array}{l} 
\tilde{\bf u}={\bf P} {\bf \Psi}_L {\bf c} \\
\tilde{\bf N}_3={\bf P} {\bf \Psi}_3 {\bf c} \\
\tilde{\bf N}_5={\bf P} {\bf \Psi}_5 {\bf c} \\ 
\tilde{\bf N}_{NL}={\bf P} {\bf \Psi}_{NL} {\bf c}
\end{array} \right. $}
\put(38,12){to$^*$}
\put(38,8){$^*$Note: one of the above}
%
\put(59,16){${\bf u} = {\bf \Phi_{L,\beta_j}}  ( {\bf P} {\bf \Phi_{L,\beta_j}})^\dagger \tilde{\bf u}$}
\put(78,16){${\bf u} = {\bf \Phi}_{L,\beta_j} {\bf a}(t)$ }
\put(65,8){$^{**}$ Note:  ${\bf \Phi}_{L,\beta_j} \,\, \text{contained in} \,\, {\bf \Psi}_L$}
\end{overpic}
\vspace*{-.6in}
\caption{Training and execution modules for the library learning and sensor location optimization with DEIM.  The training
module samples the various dynamical regimes ($\beta_1, \beta_2, \cdots, \beta_J$) through snapshots.  For each dynamical
regime, low-rank libraries are constructed for the nonlinearities of the complex system (${\bf \Phi}_{L,\beta_j}$, ${\bf \Phi}_{3,\beta_j}$, ${\bf \Phi}_{5,\beta_j}$,
${\bf \Phi}_{NL,\beta_j}$).  The DEIM algorithm is then used to select sparse sampling locations and construct the projection matrix ${\bf P}$.
The execution module uses the sampling locations to classify the dynamical regime $\beta_j$ of the complex system, reconstruct its full state (${\bf u} = {\bf \Phi_{L,\beta_j}}  ( {\bf P} {\bf \Phi_{L,\beta_j}} )^\dagger \tilde{\bf u}$), and
provide a low-rank Galerkin-POD approximation for its future (${\bf u} = {\bf \Phi}_{L,\beta_j} {\bf a}(t)$).  Note that $( {\bf P} {\bf \Phi_{L,\beta_j}} )^\dagger$ denotes the Moore-Penrose pseudo-inverse of $({\bf P} {\bf \Phi_{L,\beta_j}} )$.}
 \label{fig:deim}
\end{figure*}

\section{Libraries}\label{sec:libs}
\begin{figure}[t]
  \begin{overpic}[width=0.5\textwidth]{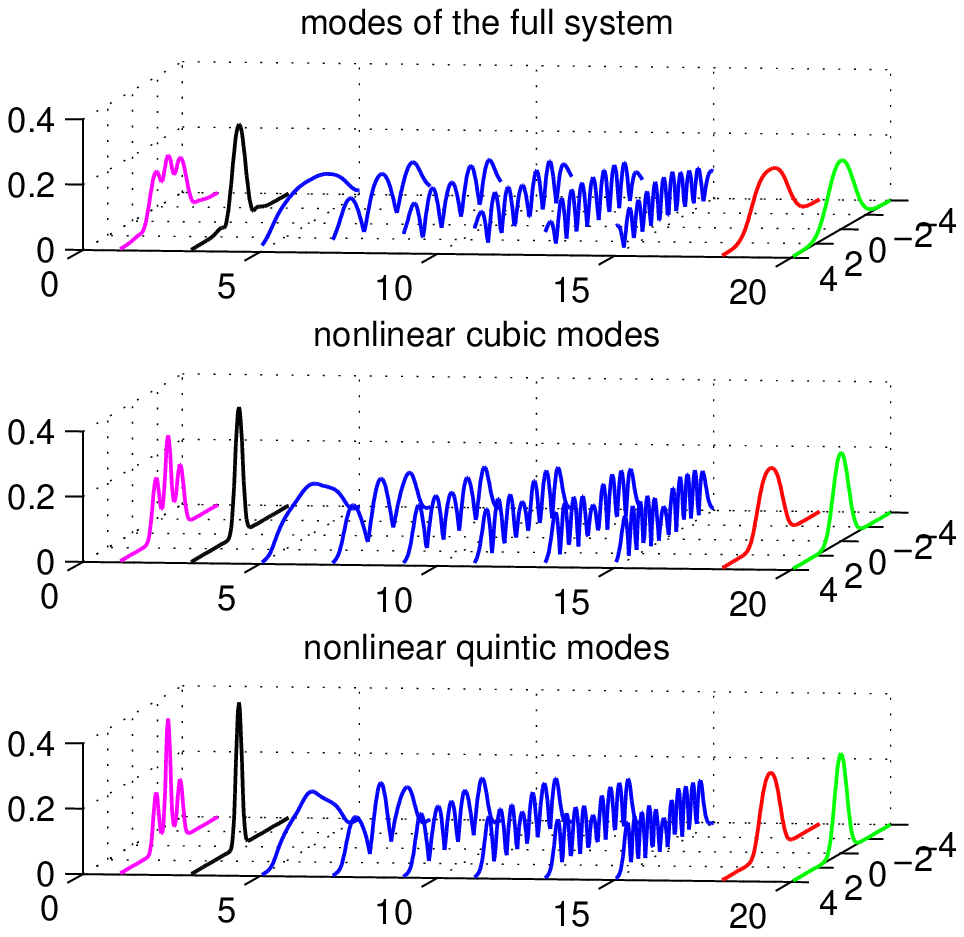}
\put(15,89){(a)}
\put(15,59){(b)}
\put(15,29){(c)}
\put(22,27){\color{magenta}{$\beta_1$}}
\put(30,27){\color{black}{$\beta_2$}}
\put(55,23){\color{blue}{$\beta_3$}}
\put(78,25){\color{red}{$\beta_5$}}
\put(85,25){\color{green}{$\beta_6$}}
\put(22,57){\color{magenta}{$\beta_1$}}
\put(30,57){\color{black}{$\beta_2$}}
\put(55,53){\color{blue}{$\beta_3$}}
\put(78,55){\color{red}{$\beta_5$}}
\put(85,55){\color{green}{$\beta_6$}}
\put(22,87){\color{magenta}{$\beta_1$}}
\put(30,87){\color{black}{$\beta_2$}}
\put(55,83){\color{blue}{$\beta_3$}}
\put(78,85){\color{red}{$\beta_5$}}
\put(85,85){\color{green}{$\beta_6$}}
\end{overpic}
\vspace*{-.4in}
\caption{Library modes for (a) the full system, (b) the cubic nonlinearity, and (c) the quintic nonlinearity. 
The modes are color coded by their dynamical regime from $\beta_1$ to $\beta_6$ as given in Table~\ref{table:vals}.
The rank-$r$ for each library is chosen by selecting the modes that comprise 99\% of the total variance for a given
dynamical regime.}
\label{figure:modes3d}
\end{figure}

As can be seen from Fig.~\ref{figure:3reg} and Table~\ref{table:vals}, generic initial conditions evolve towards
a variety of low-dimensional attractors. This suggests that each 
dynamic regime, with a given $\beta_j$, can be approximated by a small number of modes via a POD reduction. 
These modes will constitute our {\em library modes} in what follows.   For each of the six regimes $\beta_j$ in Table~\ref{table:vals},
we build a library of POD modes.  The number of POD modes $r$ is selected to capture 99\% of 
the total variance (energy). For the $\beta_1$, $\beta_2$, $\beta_5$ and $\beta_6$ regimes, only a single mode
is required so that $r=1$.  For the
 $\beta_3$ regime $r=6$, whereas for the $\beta_4$ regime, $r=14$ in order to capture the fluctuations observed. 
 Figure \ref{figure:modes3d}(a) illustrates the library POD modes in differing colors for all of the $\beta_j$ regimes except $\beta_4$.
 The exclusion of the $\beta_4$ modes in this visualization is simply due to the large number ($r=14$) necessary in
 comparison to the other dynamical regimes.  As illustrated in Fig.~\ref{fig:deim}, library building is the first step in a training 
 module aimed at {\em learning} the low-rank dynamical behavior of a complex system.

In practice, a dynamical system such as (\ref{eq:gl}) may change over time due to evolution or modulation of
the parameters $\beta_j$.  Thus the dynamics may evolve from
one attractor to another with some prescribed transition time (typically on the order of $O(1)$ time for  (\ref{eq:gl})).  
One of the primary goals of this and previous~\cite{siads,sparseSensors} work is to find {\em optimal and sparse sensor locations} 
whereby limited measurements of the system
are taken in order to classify the dynamical regime.  Interestingly, the previous efforts~\cite{siads}  used expert-in-the-loop
knowledge to help select the optimal measurement positions.  For the simple model considered here, such
expert knowledge can be acquired from familiarity with the POD library modes and considering locations of
maximal variance.  However, for a more general system, this is a difficult task that could greatly benefit
from a more principled mathematical approach.  The DEIM algorithm will provide this approach.  Moreover,
as required by DEIM, we also build low-rank libraries for the cubic and quintic terms associated with
the dynamical regimes $\beta_j$.  In doing so, we not only find nearly optimal sensor locations, but we
also circumvent the computational difficulties of the POD in evaluating the nonlinear terms.
   
To library build, consider the following linear and nonlinear functions associated with the
governing equations (\ref{eq:gl}) for a given parameter regime $\beta_j$:
\begin{subeqnarray}
\label{eq:nonlin}
 && N_{L}(U)= U \\
 && N_{3}(U)=|U|^2U \\
 && N_{5}(U)=|U|^4U \\
 && N_{NL}(U) =(i+\mu)|U|^2 U+(i\nu+\epsilon)|U|^4 U \, ,
\end{subeqnarray}
where the second and third terms are the standard cubic and quintic nonlinearities of (\ref{eq:gl})
and the last term enforces their prescribed relative weighting.  

Associated with each nonlinearity (\ref{eq:nonlin}) are a set of measurements and
snapshot matrices.  For a snapshot matrix sampled at $p$ temporal locations
$ [{\bf u}_1 \,\, {\bf u}_2 \,\,  \cdots \,\, {\bf u}_p]\in \mathbb{R}^{n\times p}$, we can construct the nonlinear $\mathbb{R}^{n\times p} $snapshot
matrices  
\begin{subeqnarray}
\label{eq:nonlin2}
&& \hspace*{-.3in} \mathbf{N}_{L}=[ {\bf u}_1 \,\, {\bf u}_2 \,\, \cdots \,\, {\bf u}_p]  \\ 
&& \hspace*{-.3in} \mathbf{N}_{3}=[ N_{3}({\bf u}_1) \,\, N_{3}({\bf u}_2) \,\, \cdots \,\, N_{3}({\bf u}_p)]  \\ 
&&  \hspace*{-.3in} \mathbf{N}_{5}=[ N_{5}({\bf u}_1) \,\,   N_{5}({\bf u}_2) \,\, \cdots \,\, N_{5}({\bf u}_p)] \\
&&  \hspace*{-.3in} \mathbf{N}_{NL}=[ N_{NL}({\bf u}_1) \,\,  N_{NL}({\bf u}_2) \,\, \cdots \,\, N_{NL}({\bf u}_p)] .
\end{subeqnarray}
The singular value decomposition of these matrices will give a basis for 
approximation of each of the nonlinearities for a given $\beta_j$ as well as the standard snapshot matrix
of POD.  Specifically, the SVD 
gives the library of modes:  ${\bf \Phi}_{L,\beta_j}$,  ${\bf \Phi}_{3,\beta_j}$, ${\bf \Phi}_{5,\beta_j}$ and ${\bf \Phi}_{NL,\beta_j}$ (See
Fig.~\ref{fig:deim}).

The POD modes can be arranged in a collection of library elements, ${\bf \Psi}_L$, ${\bf \Psi}_3$, ${\bf \Psi}_5$ or ${\bf \Psi}_{NL}$, by concatenating the POD modes from each of the different $\beta_j$ regimes.  Thus the construction of multiple libraries would take the form
\begin{subeqnarray}
\label{eq:nonlin3}
 && {\bf \Psi}_L = \left[ {\bf  \Phi}_{L,\beta_1} \,\,  {\bf \Phi}_{L,\beta_2} \,\, \cdots \,\, {\bf \Phi}_{L,\beta_6}   \right] \\
 && {\bf \Psi}_3 = \left[ {\bf  \Phi}_{3,\beta_1} \,\,  {\bf \Phi}_{3,\beta_2} \,\, \cdots \,\, {\bf \Phi}_{3,\beta_6}   \right] \\
&& {\bf \Psi}_5 = \left[ {\bf  \Phi}_{5,\beta_1} \,\,  {\bf \Phi}_{5,\beta_2} \,\, \cdots \,\, {\bf \Phi}_{5,\beta_6}   \right] \\
&& {\bf \Psi}_{NL} = \left[ {\bf  \Phi}_{NL,\beta_1} \,\,  {\bf \Phi}_{NL,\beta_2} \,\, \cdots \,\, {\bf \Phi}_{NL,\beta_6}   \right].
\end{subeqnarray}
The number of basis elements (rank) for the cubic and quintic terms in a given POD library coincides 
with the rank $r$ required for each $\beta_j$, i.e. $r=m$.   Note that the library ${\bf \Psi}_L$ is the
library containing the POD modes used for POD-Galerkin projections of the future state.  It is also
the only library constructed in previous work~\cite{siads,Bright:2013}.
Figure \ref{figure:modes3d}(b,c) shows the cubic and
quintic library modes  for (\ref{eq:gl}).  They can be compared to the standard POD modes
illustrated in Fig.~\ref{figure:modes3d}(a).   Although the modes look quite
similar, we will show that the classification can be improved with the nonlinear libraries.  Further,
evaluation of the nonlinearities through DEIM now remains a low-order computation.

\begin{figure*}[th]
\begin{overpic}[width=0.6\textwidth]{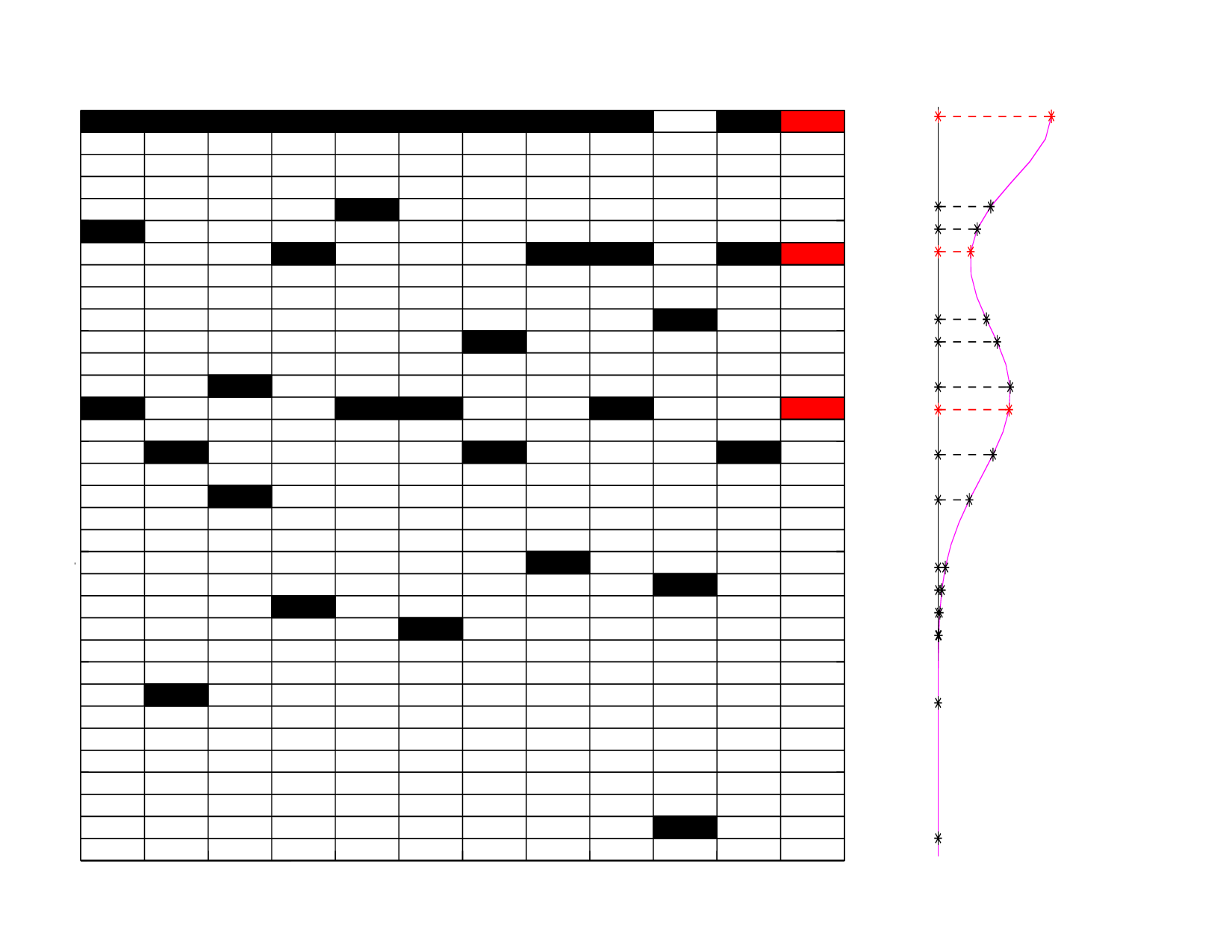}  
 \put(7,70){\color{magenta}{$\beta_1$}}
\put(12,70){\color{blue}{$\beta_3$}}
\put(17,70){\color{red}{$\beta_5$}}
\put(22,70){$\beta_{\text{all}}$}
\put(28,70){$\cdots$}
\put(0,30){\rotatebox{90}{Index $n$ of $x(n)$}}
\put(-2.5,66.5){$n\!=\!0$}
\put(0,61.5){$\vdots$}
\put(0,53.5){$\vdots$}
\put(0,15.5){$\vdots$}
\put(0,7){$\vdots$}
\put(-2.5,58){$n\!=\!5$}
\put(-3,11.5){$n\!=\!30$}
\put(6,5){\rotatebox{90}{$\begin{cases} \\ \\ \\ \\ \\  \end{cases}$}}
\put(14,2){$|U|^3$}
\put(27,5){\rotatebox{90}{$\begin{cases} \\ \\ \\ \\  \\ \end{cases}$}}
\put(35,2){$|U|^5$}
\put(48,5){\rotatebox{90}{$\begin{cases} \\ \\ \\ \\  \\ \end{cases}$}}
\put(55,2){$N(U)$}
\put(20,74){DEIM interpolation indices}
\put(78,71){${\bf \Phi}_{L,\beta_1}$}
\put(84.5,50){\rotatebox{-90}{spatial variable $x$}}
\put(83,48){\rotatebox{-90}{$\xrightarrow{\hspace*{1.5cm}}$}}
\put(83,60.5){$n\!=\!4$}
\put(81,58){$n\!=\!5$}
\put(80,55.5){$n\!=\!6$}
\put(83,51){$\vdots$}
\put(77,13){$\vdots$}
\put(78,9){$n\!=\!32$}
%
\end{overpic}
\hspace*{-.4in}
\begin{overpic}[width=0.4\textwidth]{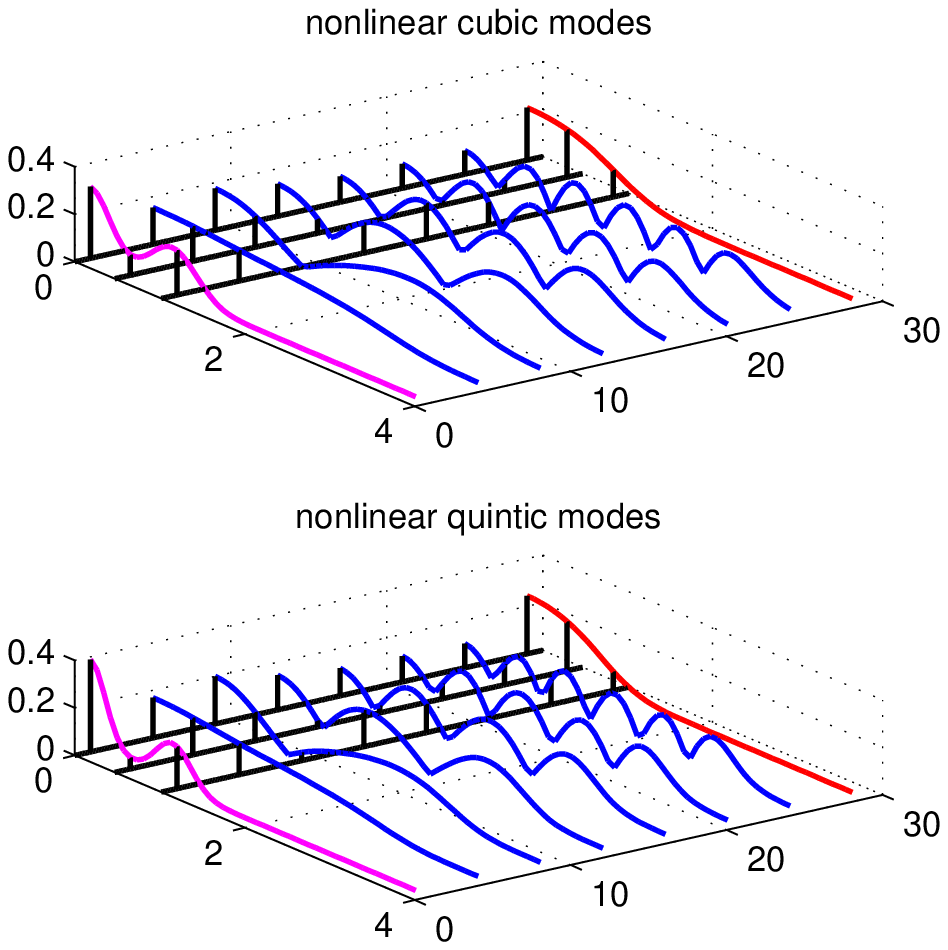}  
\put(-9,99.5){$n\!=\!0$}
\put(20,90){${\bf \Psi}_{3}$}
\put(20,42){${\bf \Psi}_{5}$}
\put(70,55){mode number}
\put(70,8){mode number}
\put(2,87){$|U|^3$}
\put(2,40){$|U|^5$}
\put(35,55){$x$}
\put(35,8){$x$}
\put(69,80){\color{red}{$\beta_5$}}
\put(69,33){\color{red}{$\beta_5$}}
\put(39,84){\color{blue}{$\beta_3$}}
\put(39,37){\color{blue}{$\beta_3$}}
\put(16,82){\color{magenta}{$\beta_1$}}
\put(16,35){\color{magenta}{$\beta_1$}}
\put(16,3){\rotatebox{80}{$\xrightarrow{\hspace*{1cm}}$}}
\put(14,1){$n\!=\!13$}
\put(10,8){\rotatebox{75}{$\xrightarrow{\hspace*{0.8cm}}$}}
\put(6,5){$n\!=\!6$}
\put(7,12){\rotatebox{70}{$\xrightarrow{\hspace*{0.6cm}}$}}
\put(0,9){$n\!=\!0$}
\put(-35,65){$\xrightarrow{\hspace*{0.15in}}$}
\put(-35,84.2){$\xrightarrow{\hspace*{0.15in}}$}
\put(-35,100.8){$\xrightarrow{\hspace*{0.15in}}$}
\end{overpic}
 \caption{Location of indices determined by DEIM for the nonlinear libraries $|U|^3$, $|U|^5$ and $N(U)$.  The spatial domain $x\in[-20,20]$ is discretized on a periodic domain with $n=1024$ points.  The center point of the domain corresponds to $x(0)=0$.  The index values are the number of grid points $n dx$ away from the center  grid point, e.g. $x(5)=5 dx$.  The left grid shows the location of the DEIM indices (black boxes) determined by the algorithm in Table~\ref{table:alg} for
 the regimes $\beta_1$, $\beta_3$ and $\beta_5$ as well as the combination of all three regimes together $\beta_{all}$.   The middle panel shows the
 library mode ${\Phi}_{L,\beta_1}$ (laid out vertically) as a function of the spatial variable $x(n)$.  Indicated on this transverse mode are the measurement 
 locations for the different DEIM nonlinearities and $\beta_j$ regimes.  The right two panels show the $\beta_1$, $\beta_3$ and $\beta_5$ modes with the black lines indicating the measurement locations for $n=0, 6$ and $13$.  This allows one to visualize where the measurement occur on the mode structures.}
 \label{figure:locs}
\end{figure*}

\section{DEIM for sensor locations}\label{sec:deim}

The idea of using a limited (sparse) number of sensors to characterize the dynamics
has previously been considered in~\cite{siads,Bright:2013,epj}.  However, no algorithm was
specified to determine the best locations for the sensors, although optimal sensor placement has been investigated in the context of categorical decisions~\cite{sparseSensors}.  Indeed, the previous
work relied on expert-in-the-loop selection of the sensors in order to classify
the dynamics.  Interestingly, the DEIM algorithm gives a principled way to 
discretely and sparsely sample the nonlinearity in order to evaluate the various
inner products for a POD reduction.   This begs the question:  would these same
DEIM spatial sampling locations make good sensor locations for classification and reconstruction?
Since the interpolation indices from the DEIM algorithm~\cite{deim} 
correspond to the entries with largest magnitude of the 
residual error between the chosen basis and its approximation at each step (see last line of the table 
\ref{table:alg}), it becomes interesting to see what the classification/reconstruction will be if we pick these 
locations for sensors.   As demonstrated in Fig.~\ref{fig:deim}, determining the sensor locations is part of a training module. 

We apply the DEIM algorithm outlined in Table~\ref{table:alg} on 
the nonlinear POD (SVD) library modes (${\bf \Psi}_3$, ${\bf \Psi}_5$ or ${\bf \Psi}_{NL}$) computed from (\ref{eq:nonlin2}) and (\ref{eq:nonlin3}).  The application
of the algorithm yields DEIM interpolation locations which we will call our
{\em sensor locations}.   Note that the indices indicate the distance away
from the center of the computational grid. Thus $x_0=0$, $x_{\pm 1}=dx$, $x_{\pm 2}=2 dx$, etc.  Or more generally,
the index $n$ corresponds to $x_n= n \, dx$.  Thus the indices depend on the specific discretization of the domain.
Sensor locations are computed for each of the nonlinearities:  ${\bf \Phi}_{3,\beta_j}$, ${\bf \Phi}_{5,\beta_j}$ and ${\bf \Phi}_{NL,\beta_j}$
for $j=1,2,3$.   Each dynamical regime $\beta_j$ and nonlinear library gives a unique set of sensor locations.  Our
goal is to evaluate the placement of 3 sensors.  Table \ref{table:senslocs} and its accompanying figure gives a vector of the indices
for the locations ${\bf x}_{\beta_j}$ of the 3 sensors found for three regimes $\beta_1$, $\beta_3$ and $\beta_5$ using the libraries
${\bf \Phi}_{3,\beta_j}$, ${\bf \Phi}_{5,\beta_j}$ and ${\bf \Phi}_{NL,\beta_j}$.
Also represented are the 3 sensor locations when all three $\beta_j$ regimes
are combined into a single library, i.e. the best sensor locations for the combined dynamic library is identified.  This
regime is represented in Table~\ref{table:senslocs} by ${\bf x}_{\beta_{\text{all}}}$.  

Application of the DEIM algorithm results in the measurement matrix ${\bf P}$ of (\ref{eq:nonlinearity}).  For 3 sensors, generically it
takes the form
\begin{equation}
\label{eq:P2}
  {\bf P} = \left[  \begin{array}{cccccccccc}  
     1 & 0          &  \cdots     &     &      & & & & \cdots  & 0 \\
     0 & \cdots   & 0             &  1 &  0  & \cdots & & & \cdots  & 0 \\
     0 & \cdots   &                &     & \cdots     & 0 & 1 & 0 &\cdots & 0
  \end{array}    \right]
\end{equation}
where the specific columns containing the nonzero entries are given by
the indices found from DEIM and shown in Table~\ref{table:senslocs}.  More precisely, this matrix is {\em exactly} the output
of the DEIM algorithm.  In our scenario, the construction of the $P$ matrix
is made for each nonlinearity as well as for each dynamical regime $\beta_j$.
This gives the nearly optimal sensor locations for the sparse sensing scheme
presented in the next section.  Figure \ref{figure:locs} illustrates the locations of
the sensors and the value of library modes at the prescribed locations for
both the cubic and quintic nonlinearities.


\begin{table}[t ]

\begin{tabular}[c]{|p{1.2cm}|p{0.4cm}|p{0.4cm}| p{0.4cm}| p{0.6cm}| p{0.4cm}|p{0.4cm}|p{0.4cm}| p{0.6cm}|p{.4cm}| p{.4cm}| p{.4cm}|p{0.6cm}| }
\hline
& \multicolumn{4}{ |c| }{Cubic} & \multicolumn{4}{ |c| }{Quintic } 
&  \multicolumn{4}{ |c| }{ Nonlinear}\\
& \multicolumn{4}{ |c| }{ $|U|^2U$} & \multicolumn{4}{ |c| }{ $|U|^4U$ } 
&  \multicolumn{4}{ |c| }{  $N(U)$}\\

\hline
Sensor
& ${\bf x}_{\beta_1}$ &${\bf x}_{\beta_3} $& ${\bf x}_{\beta_5} $&  ${\bf x}_{\beta_{all}}$ & 
 ${\bf x}_{\beta_1}$ & ${\bf x}_{\beta_3}$  &  $ {\bf x}_{\beta_5}$ &   ${\bf x}_{\beta_{all}}$ & 
 ${\bf x}_{\beta_1}$   & ${\bf x}_{\beta_3}$ & ${\bf x}_{\beta_5}$ & ${\bf x}_{\beta_{all}}$ \\
\hline
one & 0 & 0 &0  &        0 &  0 &  0 &  0  &           0 & 0 & 9 & 0 &               0\\
\hline
 two & 5 &15 & 12   &         6 &  4&13 & 10  &             6 &6 & 21 & 6   &              6  \\
\hline
three &13 &26 &17  &            22  & 13 & 23 &   15  &            20 & 13 & 32 & 15   &            13  \\
\hline
   
\end{tabular}   
\caption{Summary of sensor location vectors (indices for evaluation) from the DEIM algorithm.  The table summarizes the
findings from Fig.~\ref{figure:locs}, giving precise grid cells to be used in evaluating the nonlinear inner products in
the Galerkin-POD approximation.}
\label{table:senslocs}

\end{table}

\section{Classification}\label{sec:class}

Our goal is to make use of recent innovations in sparse sampling and compressive sensing~\cite{Donoho:2006,Donoho2,Candes:2006,Candes:2006a,Candes:2006c,Candes:2006b,Baraniuk:2007,Baraniuk:2009}
for characterizing the complex system~\cite{siads,Bright:2013,epj}.  Specifically, we wish to use a limited
number of sensors for classifying the dynamical regime of the system.  With this classification, a reconstruction
of the full state space can be accomplished and a POD-Galerkin prediction can be computed for its future.
 In general, if we have a sparse measurement $\tilde{\bf u}\in\mathbf{R}^q$, where $q$ is the number
of measurements, then 
\begin{equation}
\label{eq:P}
  \tilde{\bf u}={\bf P} 
{\bf u} \, ,
\end{equation}
where ${\bf u}$ is the full state vector and ${\bf P}$ is the sampling matrix determined by DEIM given by (\ref{eq:P2}).  
In the previous section, we constructed the matrix ${\bf P}$ for $q=3$.

The full state vector ${\bf u}$  can be approximated with the POD library modes (${\bf u}={\bf \Psi}_L {\bf c}$), therefore 
\begin{equation}
   \tilde{\bf u}={\bf P} {\bf \Psi}_L {\bf c}, 
  \label{eq:sp}
\end{equation}
where ${\bf \Psi}_L$ is the low-rank matrix whose columns are POD basis vectors concatenated across all $\beta$ regimes and ${\bf c}$ is the 
coefficient vector giving the projection of ${\bf u}$ onto these POD modes.  
If  ${\bf P}{\bf \Psi}_L$ obeys the restricted isometry property~\cite{rip} 
and ${\bf u}$ is sufficiently sparse in ${\bf \Psi}_L$, then it is possible to solve the highly-underdetermined system (\ref{eq:sp}) 
with the sparsest vector ${\bf c}$. Mathematically, this is equivalent to the optimization problem 
\begin{equation*}
 {\bf c}=\min_{{\bf c}'} ||{\bf c}'||_0, \quad \text{subject to} \quad \tilde{\bf u}={\bf P} {\bf \Psi}_L {\bf c}. 
\end{equation*}
Minimizing the $l_0$ norm is computationally an $np$-hard problem.  However,  It has been proven that 
under certain conditions, a sparse solution of equation (\ref{eq:sp}) can be found 
by minimizing the $l_1 $ norm instead~\cite{Donoho2,Candes:2006a} so that
\begin{equation}
 {\bf c}=\text{arg}\min_{{\bf c}'} ||{\bf c}'||_1, \quad \text{subject to} \quad \tilde{\bf u}={\bf P} {\bf \Psi}_L {\bf c}. 
  \label{eq:l1}
\end{equation}
The last equation can be solved through standard convex optimization methods such as  
the CVX package for Matlab.

\begin{figure}[bt]
 \begin{overpic}[width=0.4\textwidth]{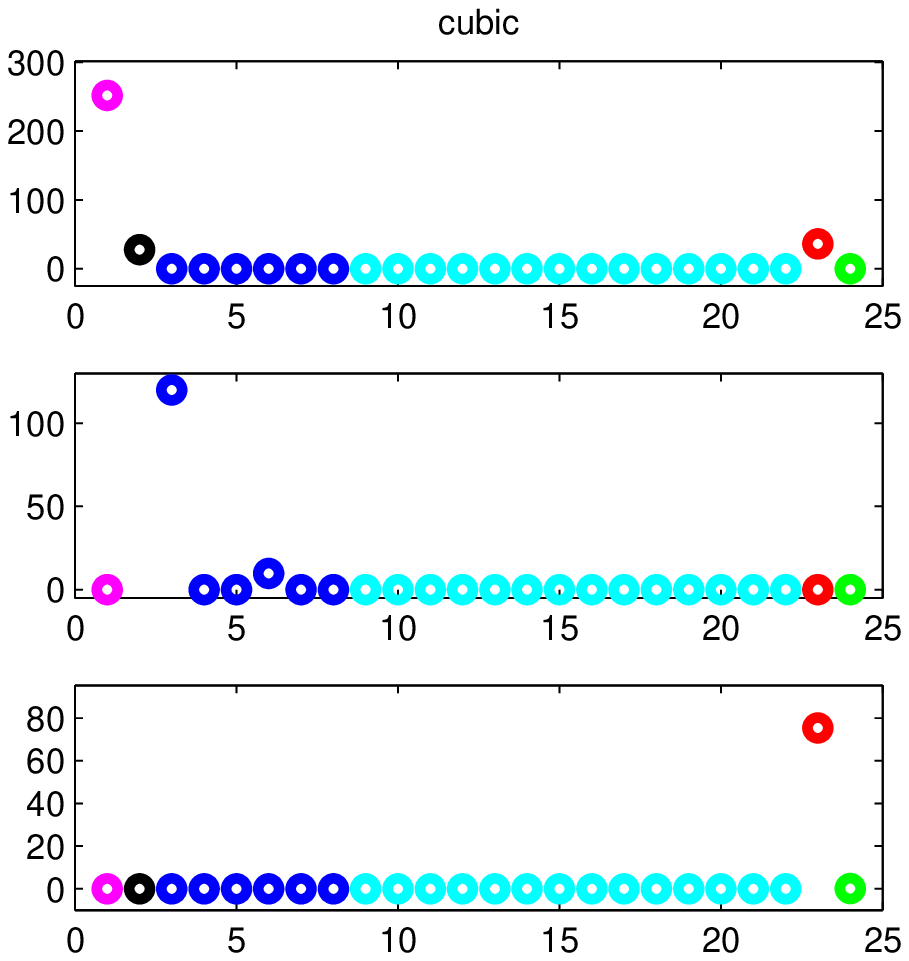}  
   \put(50, 88){$\beta_1$}
    \put(50, 58){$\beta_3$}
   \put(50, 28){$\beta_5$}
   \put(42,3){mode number $n$}
   \put(-5,25){${\bf c}_3(n)$}
   \put(-5,55){${\bf c}_3(n)$}
   \put(-5,85){${\bf c}_3(n)$}
\end{overpic}
 \caption{The values of the $24\times 1$ projection vector ${\bf c}$ from solving using a cubic measurement $\tilde{\bf u}_3=|\tilde{\bf u}|^2\tilde{\bf u}$ 
 and the cubic library ${\bf \Psi}_3$ in (\ref{eq:cj}a).  The three panels show the dominant
vector component to be in the $\beta_1$, $\beta_3$ and $\beta_5$ regime respectively, thus showing that it correctly identifies each
dynamical regime from 3 measurement locations.  The values of the colored circles correspond to the expression strength of the different library elements 
of Fig.~\ref{figure:modes3d}.}
 \label{fig:c_an}
\end{figure}

\begin{figure}[bt]
 \begin{overpic}[width=0.4\textwidth]{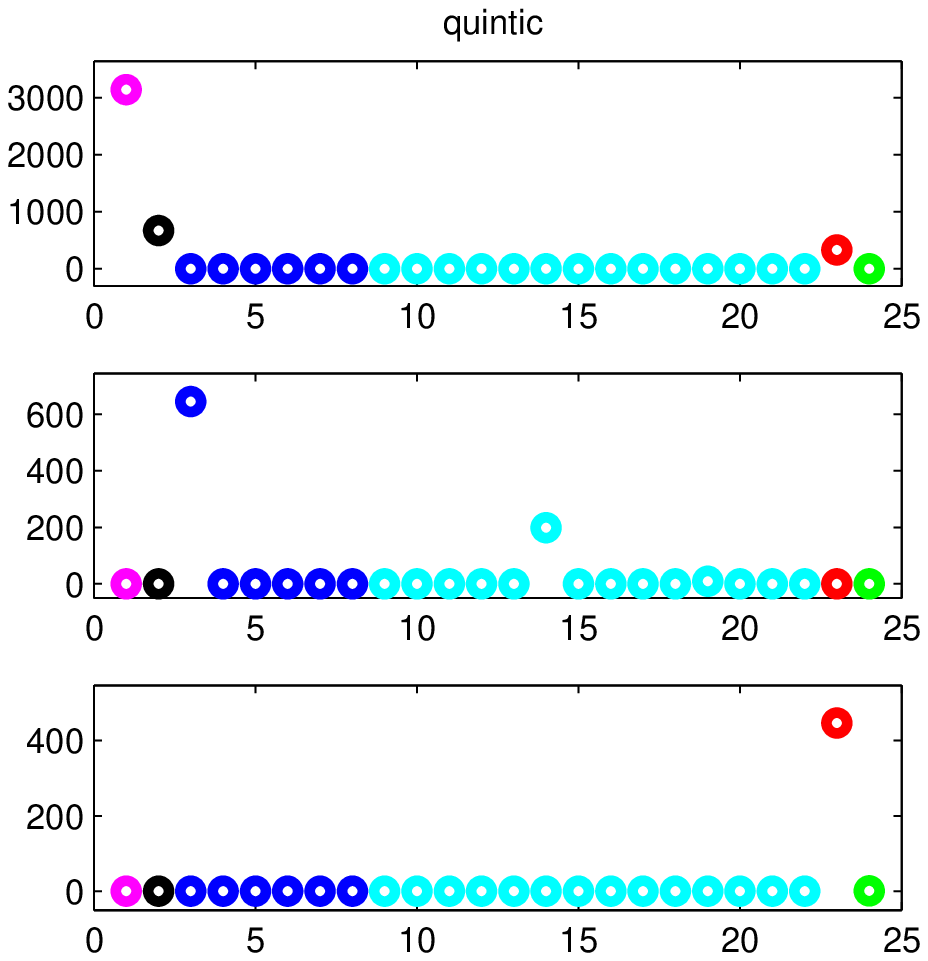}  
  \put(50, 88){$\beta_1$}
    \put(50, 58){$\beta_3$}
   \put(50, 28){$\beta_5$}
   \put(42,3){mode number $n$}
   \put(-6,25){${\bf c}_5(n)$}
   \put(-6,55){${\bf c}_5(n)$}
   \put(-6,85){${\bf c}_5(n)$}
\end{overpic}
 \caption{The values of the $24\times 1$ projection vector ${\bf c}$ from solving using a quintic measurement $\tilde{\bf u}_5=|\tilde{\bf u}|^4\tilde{\bf u}$ 
 and the quintic library ${\bf \Psi}_5$ in (\ref{eq:cj}b).  The three panels show the dominant
vector component to be in the $\beta_1$, $\beta_3$ and $\beta_5$ regime respectively, thus showing again that nonlinear measurements correctly identify 
each dynamical regime from 3 measurement locations.  The values of the colored circles correspond to the expression strength of the different library elements of Fig.~\ref{figure:modes3d}.}
 \label{figure:q_an}
\end{figure}

To classify the dynamical regime from limited measurements $\tilde{\bf u}$
(specifically 3 spatial measurements), we use the sensor locations matrix ${\bf P}$ found
from DEIM on the nonlinear libraries.  Here, the sensor locations used for ${\bf P}$ are from
all the library elements combined and the nonlinearity $N(U)$ (See the last column in Table \ref{table:senslocs} remarked
with red boxes), i.e. $n=0, 6$ and 13.   Suppose we have a linear measurement 
$\tilde{\bf u}$, then we can construct the vectors $\tilde{\bf u}_3=|\tilde{\bf u}|^2\tilde{\bf u}$ and $\tilde{\bf u}_5=|\tilde{\bf u}|^4\tilde{\bf u}$ 
and classify them using the nonlinear libraries.  Specifically, the nonlinear classification
is accomplished with:
\begin{subeqnarray}
 && \hspace*{-.5in} {\bf c}_3=\text{arg}\min_{{\bf c}_3'} ||{\bf c}_3'||_1, \quad \text{subject to} \quad \tilde{\bf u}_3={\bf P} {\bf \Psi}_3 {\bf c}_3 \\
 && \hspace*{-.5in} {\bf c}_5=\text{arg}\min_{{\bf c}_5'} ||{\bf c}_5'||_1, \quad \text{subject to} \quad \tilde{\bf u}_5={\bf P} {\bf \Psi}_5 {\bf c}_5 \, .
 \label{eq:cj}
\end{subeqnarray}
Figures~\ref{fig:c_an} and \ref{figure:q_an} show the 
coefficient vectors ${\bf c}_3$ and ${\bf c}_5$ respectively for measurements performed in the $\beta_1$, $\beta_3$ and
$\beta_5$ regimes. The vectors ${\bf c}_3$ and ${\bf c}_5$ clearly act as accurate indicator functions for the
dynamical regime.  Indeed, the DEIM algorithm for sensor location does as well
as expert-in-the-loop selections~\cite{siads,Bright:2013,epj}, but requires no extensive and pre-existing knowledge 
about the dynamical libraries.  
We can also make a categorical decision, with similar results, about the dynamical regime
the dynamics belongs to by computing error of projection onto a given library and considering which has 
the smallest error.   This is the same as sparse representation used for image classification~\cite{Wright:2009}.

The above analysis assumes that there is no noise in the measurements or the
system itself.  However, most sensors are subject to noise fluctuations which
can impact the ability of a scheme such as this to correctly identify $\beta_j$.
As a consequence, we also perform the classification task with noisy data. 
First, assume that we collect linear measurements which have additive noise. 
Denote this data by 
\begin{equation}
\label{eq:noise}
  \bar{\bf u}=\tilde{\bf u}+{\mathcal{N}}(0,\sigma^2)
\end{equation}
where $\mathcal{N}(0,\sigma^2)$ is a Gaussian distributed noise term with variance $\sigma^2$ .

%
%
%
%

\begin{table}[t ]
\begin{tabular}[c]{|p{1.8cm}|p{0.9cm}|p{0.9cm}| p{0.9cm}| p{0.9cm}| p{0.9cm}|p{0.9cm}| }
\hline
{$\beta_1$ regime }& $\beta_1$ &$\beta_2$& $\beta_3$ & $\beta_4$ & $\beta_5$ & 
$\beta_6$\\[.1in]
\hline
$|\bar{u}|^2 \bar{u}$&\bf{98.75} & 0 & 1.25 & 0 & 0 &  0 \\
\hline
$|\bar{u}|^4 \bar{u}$& \bf{91}&6.5 &2.5 & 0&0 &0  \\
\hline
$\bar{u}_3$&\bf{100}& 0& 0& 0& 0& 0\\
\hline
$\bar{u}_5$&\bf{100}& 0& 0& 0& 0& 0  \\
\hline
 && & & & &   \\
\hline 
 {$\beta_3$ regime }& $\beta_1$ &$\beta_2$& $\beta_3$ & $\beta_4$ & $\beta_5$ & 
$\beta_6$\\[.1in]
\hline
$|\bar{u}|^2 \bar{u}$ & 2.5&0 &\bf{61.75 }& 18& 17.5 & 0.25 \\
\hline
$|\bar{u}|^4 \bar{u}$&5.5 &0 &\bf{38} &34.5 &21.75 &0.25  \\
\hline
$\bar{u}_3$&0 &0 & \bf{100}&0 &0 &0 \\
\hline
$\bar{u}_5$&0 & 0&\bf{100} & 0&0 & 0\\
\hline
 && & & & &   \\
\hline
{$\beta_5$ regime} & $\beta_1$ &$\beta_2$& $\beta_3$ & $\beta_4$ & $\beta_5$ & 
$\beta_6$\\[.1in]
\hline
$|\bar{u}|^2 \bar{u}$& 5.25&0.75 &7.5 &5 & \bf{62}& 19.5 \\
\hline
$|\bar{u}|^4 \bar{u}$& 6.75&2 &6.25 & 2.5&\bf{61.25} & 21.25 \\
\hline
$\bar{u}_3$& 0&0 &0 & 0& \bf{100}&0  \\
\hline
$\bar{u}_5$& 0&0 &0 & 0& \bf{100}&0 \\
\hline

\end{tabular}   
\caption{Classification accuracy with noisy measurements $(\sigma=0.2)$ using 400 realizations in the $\beta_1$, $\beta_3$ and $\beta_5$ regimes.  
The accuracy of classification for the correct regime is denoted by the bold numbers, whereas the other percentages denote to
what extent and where misclassifications occur.
The accuracy of the classification schemes are evaluated using linear measurements ($\bar{u}$ in (\ref{eq:noise})) with the 
cubic and quintic libraries illustrated in Figs.~\ref{fig:c_an} and \ref{figure:q_an}.
  Also shown are classification results using nonlinear measurements ($\bar{\bf u}_3$ and $\bar{\bf u}_5$ in \ref{eq:noise2}).
  Nonlinear measurements, if possible, offer significant accuracy improvement and robustness to noise.}
\label{table:class}

\end{table}

In order to evaluate the classification, we need to once again compute the nonlinear terms and run the optimization algorithm
for computing the library coefficients and the associated dynamical regime.  The statistical result for 
400 trials when  $\sigma=0.2$ is shown in Table~\ref{table:class}.  One can see that the noise introduces
misclassification errors to the original 100\% accurate classification scheme.  
However, multiple measurements still give an accurate classification overall with the exception of
using the quintic library in the $\beta_3$ regime.

Interestingly, if nonlinear measurements are considered, then the results can improve drastically.
For instance, in optics, measurements are made of the intensity of the field rather than the field
itself.  This represents a simple form of a nonlinear measurement.  Thus consider the
nonlinear measurements subject to noise:
\begin{subeqnarray}
\label{eq:noise2}
 && \bar{\bf u}_3=|\tilde{\bf u}|^2\tilde{\bf u}+\mathcal{N}(0,\sigma^2) \\ 
&& \bar{\bf u}_5=|\tilde{\bf u}|^4\tilde{\bf u} +\mathcal{N}(0,\sigma^2) \, .
\end{subeqnarray}
The classification results for this case are also shown in Table~\ref{table:class}.  Note
the clear improvement (100\% accuracy) in using nonlinear measurements for classification tasks.
Thus if the noise is driven by the sensor itself, then nonlinear measurements may
be quite advantageous.

\section{Reconstruction and the Galerkin-POD Approximation}\label{sec:recon}

The classification step of the last section identifies the dynamical regime of the complex system
by using sparsity promoting $\ell_1$ optimization on the learned libraries.  Once the correct $\beta_j$
regime is determined, reconstruction of the solution and a future state prediction can be achieved
through the POD-Galerkin approximation.  Specifically, once the dynamical regime $\beta_j$ has
been identified, then a subset of modes $\mathbf{\Psi}_L \rightarrow \mathbf{\Phi}_{L,\beta_j}$ form the correct modal basis for a POD-Galerkin approximation.

To be more precise, recall that only a limited number of measurements are made as in (\ref{eq:P}).  But
now ${\bf u}=\mathbf{\Phi}_{L,\beta_j} {\bf c}$ where the vector ${\bf c}$ is now the projection onto the smaller
set of library modes associated with a single $\beta_j$.  Thus instead of (\ref{eq:sp}), we now we have
\begin{equation}
\label{eq:sp2}
  \tilde{\bf u} = {\bf P} \mathbf{\Phi}_{L,\beta_j} {\bf c} \, .
\end{equation}
 Unlike the classification step, we can now determine ${\bf c}$ by simply solving the above equation
 using a standard Moore-Penrose pseudo-inverse operator $\dagger$~\cite{tre} so that ${\bf c}= ({\bf P}\mathbf{\Phi}_{L,\beta_j})^\dagger \tilde{\bf u}$, i.e. it
 solves for ${\bf c}$ by minimizing the $\ell_2$ norm.
With ${\bf c}$ determined, the reconstruction of the solution thus follows:
\begin{equation}
  {\bf u}= {\bf \Phi}_{L,\beta_j} ({\bf P}{\bf \Phi}_{L,\beta_j})^\dagger \tilde{\bf u} 
  \label{eq:reconstruct}
\end{equation}
This is the reconstruction of the system given the sparse measurement vector $\tilde{\bf u}$ and a classification $\beta_j$.
The POD-Galerkin approximation for the future state can then be accomplished by using (\ref{eq:pod}) and with the 
DEIM algorithm for evaluating the nonlinearities (\ref{eq:nonlinearity}).  The initial condition for the POD-Galerkin is given
from (\ref{eq:reconstruct}).  Thus as advocated in previous work~\cite{siads,Bright:2013}, accurate classification is accomplished with $\ell_1$ optimization (decoding)
while the more standard $\ell_2$ norm is used for reconstruction and POD-Galerkin projection (encoding).
Figure~\ref{fig:deim} illustrates the execution state outlined here for classification, reconstruction and projection.

\section{Conclusions and Outlook}\label{sec:conclusion}

In conclusion, we advocate a general theoretical framework for complex systems whereby low-rank libraries representing
the optimal modal basis are constructed, or learned, from snapshot sampling of the dynamics.  In order to make
model reduction methods such as POD computationally efficient, especially in evaluating the nonlinear terms 
of the governing equations, nonlinear libraries are also constructed during the learning stage.  This allows for
the application of the discrete empirical interpolation method which identifies a limited number of spatial sampling
locations that can allow for reconstruction of the nonlinear terms in a low-dimensional manner.  Such sparse
sampling of the nonlinearity is directly related to compressive sensing strategies whereby a small number of
sensors can be used to characterize the dynamics of the complex system.  Indeed, the POD method, when
combined with DEIM and compressive sensing, can (i) correctly identifying the dynamical parameter regime,
(ii) reconstruct the full state dynamics and (iii) produce a low-rank prediction of the future state of the complex
system. All of these tasks are accomplished in a low-dimensional way, unlike standard POD-Galerkin models
whose nonlinearities can prove to be computationally inefficient.

To be more precise about our learning algorithm for the complex system, We construct
the library modes representing the dynamics by the $\ell_2$-optimal proper orthogonal decomposition.  Several
libraries are constructed:  one for linear snapshot measurements, one for each nonlinear term, and one which
combines all the nonlinear terms together with their prescribed weightings.  The DEIM algorithm then
allows us to identify sparse measurement locations capable of both classifying the dynamics regime of
the complex system and efficiently evaluating the nonlinear inner products for a POD-Galerkin projection of the system.
Indeed, the dynamical state is identified from limited noisy measurements using the sparsity promoting $\ell_1$ norm and the compressive sensing architecture.  The strategy for building modal libraries by concatenating truncated POD libraries across a range of relevant bifurcation parameters may be viewed as a simple machine learning implementation.  
The resulting modal libraries are a natural sparse basis for the application of compressive sensing.  After the expensive one-time library-building procedure, accurate identification, projection, and reconstruction may be performed entirely in a low-dimensional framework.

With three DEIM determined sensor locations, it is possible to accurately classify bifurcation regimes, reconstruct the low-dimensional content, and simulate the Galerkin projected dynamics of the complex Ginzburg Landau equation.  
In addition, we investigate the performance of sparse representation with the addition of sensor noise.  
For moderate noise levels, the method accurately classifies the correct dynamic regime.  Nonlinear measurements dramatically improve the classification procedure.  
Interestingly, the DIEMs algorithm not only provides nearly optimal sensor positioning, it also helps perform
POD-Galerking truncations in a fully low-rank manner, thus avoiding the computational expense of evaluating nonlinear
terms using the POD methodology.
Overall, the combination of $\ell_2$ low-rank representations and $\ell_1$ sparse sampling enables efficient characterization and manipulation of low-rank dynamical systems.  

For modern complex systems, it is known that nonlinearity plays
a dominant role and shapes the underlying spatio-temporal dynamics and modal structures, 
thus necessitating a new approach, such as that presented here, for extracting these 
critical structures.  As has been demonstrated, although nonlinearity drives new modal structures, it does not destroy the underlying
low-dimensional nature of the dynamics.  Methods that take advantage of such underlying structure are critical for 
developing theoretical understanding and garnering insight into the fundamental interactions of a vast array of physical, engineering and biological systems.

\section*{Acknowledgements}

We are grateful for discussions with Ido Bright, Bingni W. Brunton, Xing Fu, Josh Proctor and Jonathan Tu.  J. N. Kutz acknowledges support
from the U.S. Air Force Office of Scientific Research (FA9550-09-0174).

\end{document}